\UseRawInputEncoding
\documentclass[prd,twocolumn,showpacs,amsmath,amssymb]{revtex4}
\pdfoutput=1
\usepackage{amssymb}
\usepackage{mathrsfs}
\usepackage{txfonts}
\usepackage{hyperref}
\usepackage{graphicx}
\usepackage{dcolumn}
\usepackage{bm}
\usepackage{xcolor}
\usepackage{soul}

\begin{document}

\title{Primordial non-Gaussianity in the warm $k$-inflation}

\author{Chao-Qun Shen\textsuperscript{1}}
\author{Xiao-Min Zhang\textsuperscript{1}}
\thanks{Corresponding author}
\email{zhangxm@mail.bnu.edu.cn}
\author{Zhi-Peng Peng\textsuperscript{2} }
\email{zhipeng@mail.bnu.edu.cn}
\author{He Liu\textsuperscript{1}}
\email{liuhe@qut.edu.cn}
\author{Xi-Bin Li\textsuperscript{3}}
\email{lxbbnu@mail.bnu.edu.cn}
\author{Peng-Cheng Chu\textsuperscript{1}}
\email{kyois@126.com}

\affiliation{\textsuperscript{1}School of Science, Qingdao University of Technology, Qingdao 266033, China\\ \textsuperscript{2}College of Science, Henan University of Technology, Zhengzhou 450001, China \\ \textsuperscript{3}College of Physics and Electronic Information, Inner Mongolia Normal University, 81 Zhaowuda Road, Hohhot, 010022, Inner Mongolia, China}

\begin{abstract}
This paper presents and investigates non-Gaussian perturbations for the warm $k$-inflation model that is driven by pure kinetic energy. The two complementary components of the overall non-Gaussianity are the three-point and four-point correlations. The intrinsic non-Gaussian component, denoted as the nonlinear parameter $ f_{NL}^{int} $, is rooted in the three-point correlation for the inflaton field. Meanwhile, the $\delta N$ part non-Gaussianity, denoted as $ f_{NL}^{\delta N} $, is the contribution attributed to the four-point correlation function of the inflaton field. In this paper, the above two components in warm k-inflation are individually computed and analyzed. Then, comparisons and discussions between them are conducted, and the non-Gaussian theoretical results are compared with experimental observations to determine the range of model parameters within the allowable range of observation.
\end{abstract}
\maketitle

\section{\label{sec1}Introduction}
The inflationary model is the most appealing for explaining issues of the standard cosmology model like horizon, flatness, and monopole puzzles. The dominant inflationary models have been categorized into two paradigms via the previous study. In the first paradigm, which is called cold inflation, the inflationary field diminishes potential and swiftly drives the universe toward the supercooling phase. To solve the ``graceful exit" problem, there must be a reheating period to bring the universe back to a radiation-dominated phase at the end of inflation \cite{Guth1981,Linde1982,Albrecht1982}. Warm inflation is the other paradigm, and in this paradigm, there is no reheating phase during warm inflation because radiation is continuously produced by interactions of the inflaton field with several other subdominant boson or fermion fields. The universal expansion exits gracefully, and the radiation energy density becomes dominant smoothly when the inflation ends \cite{Berera1995,Fang1995,Bardeen1983,Berera1999}.

The most general inflationary scenarios are based on the potential energy for the scalar field, in which the potential energy outweighs the dynamic energy and causes the universe to grow quasi-exponentially. However, Mukhanov first proposed the "$k$-inflation" model that is driven by kinetic energy terms for a scalar field $\phi$ \cite{Armendariz-Picon1999}. In string theory, nonstandard kinetic components are studied based on the existence of higher-order corrections to the effective action of the scalar field. So, the $k$-inflation picture introduces novel mechanics to the inflation model. The $k$-inflation model has been generalized to warm inflation in our previous work \cite{Peng2016,Peng2018}. The standard potential-driven warm inflation theory has been extended to warm $k$-inflationary case including cosmological perturbations \cite{Peng2018}. In addition, there are some kinds of new and more effective warm inflationary theory proposed recently \cite{Bastero-Gil2016,Bastero-Gil2021}. Warm inflationary scenario can have interesting features to construct an unifying picture of very early inflation with dark matter or dark energy \cite{Rosa2019,Ventura2019,ZhangGuo2021}. From many different warm inflationary cases, it can be concluded that there are large differences among different theory of warm inflation perturbations, which incorporates strong and weak regimes \cite{Bastero-Gil2014,Moss2008,Peter2009}.

Calculating the two-point correlation function is a method to distinguish different inflation models, where two-point correlation statistical information is reflected by the power spectrum. However, the statistical information is limited in the power spectrum, so it cannot distinguish these models more effectively. Therefore, bispectral and non-Gaussian measurements that distinguish various inflation models become necessary and receive much attention \cite{Tegmark2004,Jeong2009}. When analyzing inflation models, non-Gaussianity is usually an important consideration. The Gaussian term is dominated in inflationary perturbations, i.e., the dominant term for slow roll inflation fluctuations deviates marginally from the pure Gaussian term. Gaussianity dominates the primordial curvature perturbations \cite{Heavens1998,AR2014}. The three-point function and its Fourier transform, i.e., the bispectral representation has the leading statistics ability to differentiate between non-Gaussian and Gaussian perturbations. The topic of non-Gaussianity in warm inflation has been studied in several scenarios, such as a general type of noncanonical, and strong or weak dissipative regimes \cite{Moss2007,Gupta2002,Zhang2019}. In recent years, numerous studies have investigated the primordial non-Gaussianity generated by warm inflationary models, but the research on non-Gaussianity in the warm k-inflationary case is still blank, which is the aim of this paper.

This paper provides the theory of non-Gaussianity in warm $k$-inflation. First, the dynamical equations of warm k-inflation for the flat Friedmann-Robertson-Walker (FRW) background are introduced. Typically, the observation limit of the nonlinear parameter $ {f_{NL}} $ is established to estimate the non-Gaussian level. Second, two complementary parts of the non-Gaussianity are discussed. The first one is the three-point correlation, which is presented from the field self-interaction and can be calculated by solving slow roll perturbation equations. The other one is the four-point correlation function, which can be derived using the $\delta N$ formalism. Particularly, in the case of multiple inflation, the non-Gaussianity can be calculated conveniently using the $\delta N$ form. In this paper, the $\delta N$ formalism is first introduced for calculating non-Gaussianity in warm $k$-inflation, and then the $\delta N$ part non-Gaussianity is computed.

The rest of this paper is organized as follows: In Section II, the warm $k$-inflation model and its fundamental equations are introduced. In Section III, the $\delta N$ formalism is applied to compute the non-Gaussianity produced by the four-point correlation \cite{Sasaki2008,Allen2006}, and the non-Gaussianity in the three-point correlation function is computed. Finally, in Section IV, the total results and discussions are obtained.

\section{\label{sec2}THE FRAMEWORK OF WARM K-INFLATION}
Since the universe was built from multiple components during warm inflation, the total matter action can be represented as
\begin{equation}\label{eq1}
S = \int {{d^4}x\sqrt { - g} \left[ \mathcal{L}\left( {X,\phi } \right) + { \mathcal{L} _\gamma } + { \mathcal{L} _{{\mathop{\rm int}} }}\right]},
\end{equation}
where $g$ denotes the metric determinant, ${\mathcal{L} \left( {X,\phi } \right)}\ $ denotes the Lagrangian density in the inflaton field, $ \mathcal{L} _\gamma $ donates the Lagrangian density in the radiation field, and $ {{\cal L}_{{\mathop{\rm int}} }} $ describes the inflaton field interacting with other fields.

The warm $k$-inflation model that was employed originally is
\begin{equation}\label{eq2}
	{\cal L} = K\left( \phi  \right)X + L\left( \phi  \right){X^2} +  \cdots.
\end{equation}
By redefining
\begin{equation}\label{eq3}
	{\phi _{new}} = \int {d{\phi _{old}}{{\cal L}^{\frac{1}{4}}}\left( {{\phi _{old}}} \right)},
\end{equation}
we can rewrite the Lagrangian density more simply and concisely as
\begin{equation}\label{eq4}
	{\cal L} = K\left( {{\phi _{new}}} \right)X_{new} + X_{new}^2.
\end{equation}
In this paper, the new field is used without the subscript "new" for convenience.

Due to the redefined $\phi$ in our pure kinetic warm inflationary model, $\phi$ does not have the usual dimension of mass, and the major parameters in the model lack the traditional dimension of the canonical warm inflation. The inflaton field is dimensionless, and the dimensions of the corresponding major parameters are given below:
\begin{eqnarray}\label{eq5}
&&\left[ X \right] = {\left[ m \right]^2},\left[ {\dot \phi } \right] = \left[ m \right],\left[ {K\left( \phi  \right)} \right] = {\left[ m \right]^2},\nonumber\\
&&\Gamma  = {\left[ m \right]^3},  \left[ r \right] = {\left[ m \right]^2}.
\end{eqnarray}

Then, the Lagrangian density in the pure kinetic inflation field can be simply expressed as in \cite{Armendariz-Picon1999,Peng2016}:
\begin{equation}\label{eq6}
\mathcal{L} \left( {X,\phi } \right) = K\left( \phi  \right)X + {X^2},
\end{equation}
where $ X = \frac{1}{2}{g^{\mu \nu }}{\partial _\mu }\phi {\partial _\nu }\phi $ and $ K\left( \phi  \right) $ is called the ``kinetic function", which is the function of the inflaton field $ \phi $. A fluid with the energy-momentum tensor can appropriately characterize the inflaton field in an FRW universe with a flat spatial structure:
\begin{equation}\label{eq7}
	T_{\mu \nu }^\phi  = \left( {{\rho _\phi } + {p_\phi }} \right){u_\mu }{u_\nu } - {p_\phi }{g_{\mu \nu }},
\end{equation}
where $ \rho _\phi $, $ p_\phi $ and $ u_\mu $, and $ u_\nu $ denote the energy density, pressure, and four-velocity for the inflaton field, respectively. The energy momentum of the inflaton field is determined by varying the action for the inflaton field relative to the metric,
\begin{small}
\begin{equation}\label{eq8}
T_{\mu \nu }^\phi  = \frac{2}{{\sqrt { - g} }}\frac{\delta S_{\phi}}{{\delta {g^{\mu \nu }}}} = {{\cal L}_X}{\partial _\mu }\phi {\partial _\nu }\phi  - {g_{\mu \nu }}{\cal L}\left( {X,\phi } \right),
\end{equation}
\end{small}
where $ \mathcal{L} _X $ represents a Lagrangian partial derivative of $ \mathcal{L} \left( {X,\phi } \right) $ with respect to X. Then we have:
\begin{equation}\label{eq9}
{\rho _\phi } =2X\mathcal{L}_X-\mathcal{L}= K\left( \phi  \right)X + 3{X^2},
\end{equation}
\begin{equation}\label{eq10}
{p_\phi } =\mathcal{L}= K\left( \phi  \right)X + {X^2},
\end{equation}
and
\begin{equation}\label{eq11}
	{u_\mu } = \sigma \frac{{{\nabla _\mu }\phi }}{{\sqrt {2X} }},
\end{equation}	
where $ \sigma $ denotes the symbol of $ {\dot \phi } $.

In the warm $k$-inflation, the major dynamical equations are given by \cite{Peng2016}:
\begin{equation}\label{eq12}
{H^2} = \frac{1}{{3M_P^2}}\left( {{\rho _\phi } + {\rho _\gamma }} \right) = \frac{1}{{3M_P^2}}\rho,
\end{equation}	
\begin{equation}\label{eq13}
{{\dot \rho }_\phi } + 3H\left( {{\rho _\phi } + {p_\phi }} \right) =  - \Gamma {{\dot \phi }^2},
\end{equation}
\begin{equation}\label{eq14}
{{\dot \rho }_\gamma } + 4H{\rho _\gamma } = \Gamma {{\dot \phi }^2},
\end{equation}
where $ M_p^2 \equiv {\left( {8\pi G} \right)^{ - 1}} $, $ \rho $ is the total energy density of the universe, ${{\rho _\gamma }}$ denotes the energy density for radiation, and $\Gamma$ denotes the dissipative coefficient for warm inflation. Since the inflaton field and radiation are the dominant components of the universe during inflation, the total energy density $\rho$ and pressure $p$ are represented as
\begin{equation}\label{eq15}
	\rho  = K\left( \phi  \right)X + 3{X^2} + {\rho _\gamma },
\end{equation}	
and
\begin{equation}\label{eq16}
	 p = K\left( \phi  \right)X + {X^2} + \frac{1}{3}{\rho _\gamma }.
\end{equation}	
This paper considers a homogeneous background scalar field, so we have $ X = \frac{1}{2}\mathop{\dot \phi } $. The motion equation for the inflaton field can be determined by varying the action function and considering the thermal damped effect in the warm inflationary scenario:
 \begin{equation}\label{eq17}
 	\left( {3{{\dot \phi }^2} + K} \right)\ddot \phi  + 3H\left( {{{\dot \phi }^2} + K + r} \right)\dot \phi  + \frac{1}{2}{K_\phi }\dot \phi^2  = 0,
 \end{equation}
where $ K_{\phi} $ is a derivative of $ \phi $, and the dissipative strength parameter is defined as $r = \frac{\Gamma }{3H}$.

Due to the difficulty of solving the exact model through Eqs. (\ref{eq12}), (\ref{eq14}), and (\ref{eq17}), a slow roll approximation is frequently used. A stability analysis was conducted to confirm the slow roll situation for the dynamical systems staying near the quasi-exponential inflationary attractor for a significant number of Hubble times\cite{Peng2016}. The conditions of a slow roll are:
\begin{equation}\label{eq18}
\epsilon\ll 1, \left| \eta  \right| \ll \frac{{{{\cal L}_X}}}{{\left( {{{\cal L}_X} + r} \right)c_s^2}},\left| b \right| \ll 1, \left| c \right| < 4, \frac{{rc_s^2}}{{{{\cal L}_X}}} \ll 1 - 2c_s^2,
\end{equation}
where the parameters in the above equations are defined as
\begin{eqnarray}\label{eq19}
&&\epsilon= \frac{{{K_\phi }\dot \phi }}{{HK}},\eta = \frac{{{K_{\phi \phi }}\dot \phi }}{{HK_{\phi}}},b = \frac{{{\Gamma _\phi }\dot \phi }}{{H\Gamma }},c=\frac{{T{\Gamma _T}}}{\Gamma },\nonumber\\
&&c_s^2\simeq \frac{\dot p}{\dot \rho } = \frac{{{\dot \phi }^2} + K}{{3{{\dot \phi }^2} + K}} {\rm{ < }}1.
\end{eqnarray}
In the parameter definitions, the subscripts represent the partial derivative of the quantities for the inflaton field or temperature, while the dot represents the time derivative of quantities. Besides, quasi-exponential warm $k$-inflation proves that the term $ {{\dot \phi }^2} + K + r $ represents a small positive quantity, and $ {{\dot \phi }^2} $ and $ \left| {K\left( \phi  \right)} \right|$ have the same order \cite{Peng2016}. Then, the energy density of the inflaton field is on the order of $ \frac{1}{4}\left| K \right|{{\dot \phi }^2} $, i.e., $ {\rho _\phi } \sim \frac{1}{4}\left| K \right|{{\dot \phi }^2}$, and the inflationary period now enters the slow roll phase and has surpassed the radiation period, i.e., $ {\rho _\gamma }{\rm{ < }}{\rho _\phi } $. Thus, the Friedmann equation Eq. (\ref{eq12}) can be reduced to
\begin{equation}\label{eq20}
	{H^2} \simeq \frac{1}{{3M_P^2}}\left( {\frac{1}{4}\left| K \right|{{\dot \phi }^2}} \right).
\end{equation}
Then, based on the slow roll approximations and guaranteed by the slow roll conditions during the inflation \cite{Weinberg2008,Peng2016}, Eq. (\ref{eq17}) can be rewritten as
\begin{equation}\label{eq21}
6H\left( {{{\dot \phi }^2} + K + r} \right) \simeq  - {K_\phi }\dot \phi,
\end{equation}
where the dissipative strength of the model is determined by the dissipative strength parameter $r$. In the weak dissipation regime ($r \ll 1 $), the background dynamical evolution in the inflaton field is not affected by dissipation because it is too weak. However, the field fluctuations will be modified by the thermal variations in the radiation energy density, which will also affect the primordial spectrum of perturbations. In the strong dissipation regime ($ r \gg 1 $), the background dynamics and fluctuations will be dominated by thermal dissipation, making it simpler to satisfy slow roll conditions.

Generally, this paper considers radiation production to be quasi-stable, i.e., ${{\dot \rho }_\gamma } \ll 4H{\rho _\gamma }$. The density of radiation represented thus is
\begin{equation}\label{eq22}
{\rho _\gamma } = \kappa {T^4} \simeq \frac{3}{4}r{{\dot \phi }^2}.
\end{equation}
Based on Eqs. (\ref{eq19}), (\ref{eq20}), and (\ref{eq21}), the relationship between $ {\rho _\gamma } $ and $ {\rho _\phi } $ is obtained:
\begin{equation}\label{eq23}
	{\rho _\gamma } = \frac{r}{{2\left( {{{\cal L}_X} + r} \right)}}\epsilon{\rho _\phi },
\end{equation}
where $ {{\cal L}_X} = {{\dot \phi }^2} + K $. The following condition explains the epoch during which warm $k$-inflation takes place,
\begin{equation}\label{eq24}
	{\rho _\phi } \gg \frac{{2\left( {{{\cal L}_X} + r} \right)}}{r}{\rho _\gamma }.
\end{equation}
In contrast, inflation ceases when the universe reaches a phase dominated by radiation, and this occurs when $ \epsilon \simeq 1 $, indicating $ {\rho _\phi } \simeq \frac{{2\left( {{{\cal L}_X} + r} \right)}}{r}{\rho _\gamma } $ at the end of inflation. The number of e-folds of inflation is given by
\begin{equation}\label{eq25}
	N = \int_{{t_i}}^{{t_e}} {Hdt = \int_{{\phi _i}}^{{\phi _e}} {\frac{H}{{\dot \phi }}} } d\phi  \simeq \frac{\sigma }{{2\sqrt 3 {M_p}}}\int_{{\phi _i}}^{{\phi _e}} {\sqrt { - K\left( \phi  \right)} } d\phi,
\end{equation}
where $ {{\phi _i}} $ is the initial value of the inflaton field, and $ {{\phi _e}} $ is the final value.

\section{\label{sec3}THE NON-GAUSSIANITY IN WARM $K$-INFLATION}
Non-Gaussianity in warm $k$-inflation is comprised of two complementary elements: the $\delta N$ component and the intrinsic component. These two components are now determined separately.
\subsection{\label{sec31}the $\delta N$ part non-Gaussianity}

The $\delta N$ formalism is often used to compute the non-Gaussian property of multi-field inflation, which can be found in numerous works \cite{Lyth2005,Rodr¨ªguez2005,Sasaki1998}. According to cosmological observations, the primordial curvature perturbation, denoted as $ \zeta $, is a Gaussian dominated term with a nearly scale-invariant spectrum.

The expansion $N(t,\mathbf{x})\equiv\ln\left[\frac{\tilde{a}(t)}{a(t_{in})}\right]$ from any beginning flat slice at time $t_{in}$ to a final slice can be described with uniform energy density, where $\tilde{a(t,\mathbf{x})}$ is the locally-defined scale factor. As $\delta N$ formalism suggests \cite{Lyth2005,Boubekeur2006,Rodr¨ªguez2005}, and considering that the curvature perturbation $ \zeta $ is almost Gaussian, we have:
\begin{equation}\label{eq26}
\zeta \left( {t,{\mathbf{x}}} \right) \simeq \delta N = N\left( {t,{t_{\rm{i}}},{\rm{x}}} \right) - N\left( {t,{t_i}} \right).
\end{equation}
For good accuracy, $\delta N$ can perform series expansion of the initial scalar field,
\begin{equation}\label{eq27}
	\delta N = {N_{,I}}\delta {\phi ^I} + \frac{1}{2}{N_{,IJ}}\delta {\phi ^I}\delta {\phi ^J} + \cdots,
\end{equation}
where $N_{,I}\equiv\frac{\partial N}{\partial\phi^I}$, and $N_{,IJ}\equiv\frac{\partial^2 N}{\partial\phi^I\partial\phi^J}$.
In the equation above, the items above the second order are omitted. Finally, the two-point correlation function and three-point correlation function could be stated in the form of $ \delta N $:
\begin{equation}\label{eq28}
	{\mathcal{P}_\zeta } = {\delta ^{IJ}}{N_{,I}}{N_{,J}}\mathcal{P}_{\phi\ast},
\end{equation}
and
\begin{eqnarray}\label{eq29}
 &&\left\langle {\zeta \left( {{{\pmb{k}}_{\rm{1}}}} \right)\zeta \left( {{{\pmb{k}}_{\rm{2}}}} \right)\zeta \left( {{{\pmb{k}}_{\rm{3}}}} \right)} \right\rangle\nonumber\\
	 &&={N_{,I}}{N_{,J}}{N_{,K}}\left\langle {\delta {\phi ^I}\left( {{{\pmb{k}}_{\rm{1}}}} \right)\delta {\phi ^J}\left( {{{\pmb{k}}_{\rm{2}}}} \right)\delta {\phi ^K}\left( {{{\pmb{k}}_{\rm{3}}}} \right)} \right\rangle \nonumber\\
	&& +\frac{1}{2}{N_{,I}}{N_{,J}}{N_{,KL}}\left\langle {\delta {\phi ^I}\left( {{{\pmb{k}}_{\rm{1}}}} \right)\delta {\phi ^J}\left( {{{\pmb{k}}_{\rm{2}}}} \right)\left( {\delta {\phi ^K} \star \delta {\phi ^L}} \right)\left( {{{\pmb{k}}_{\rm{3}}}} \right)} \right\rangle \nonumber\\
&&+ perms,
\end{eqnarray}
where $\star$ represents convolution, and the expanded high-order term is not written down.
Now, this paper introduces the nonlinear parameter $ {f_{NL}} $ describing the non-Gaussian level, and they stand for observational limits. The power spectrum and bispectrum for curvature perturbation are defined as
\begin{equation}\label{eq30}
	\left\langle {\zeta \left( {{{\pmb{k}}_{\rm{1}}}} \right)\zeta \left( {{{\pmb{k}}_{\rm{2}}}} \right)} \right\rangle  \equiv {\left( {2\pi } \right)^3}\delta^3 \left( {{{\pmb{k}}_{\rm{1}}} + {{\pmb{k}}_{\rm{2}}}} \right)\frac{{2{\pi ^2}}}{{k_1^3}}{\mathcal{P}_\zeta }\left( {{k_1}} \right),
\end{equation}
and
\begin{equation}\label{eq31}
	\left\langle {\zeta \left( {{{\pmb{k}}_{\rm{1}}}} \right)\zeta \left( {{{\pmb{k}}_{\rm{2}}}} \right)\zeta \left( {{{\pmb{k}}_{\rm{1}}}} \right)} \right\rangle  \equiv {\left( {2\pi } \right)^3}\delta^3 \left( {{{\pmb{k}}_{\rm{1}}} + {{\pmb{k}}_{\rm{2}}} + {{\pmb{k}}_{\rm{3}}}} \right){B_\zeta }\left( {{k_1},{k_2},{k_3}} \right),
\end{equation}
where $\mathcal{P}_{\zeta}(k)\equiv\frac{k^3}{2\pi^2}P_{\zeta}(k)$.

The bispectrum can be expressed as
\begin{equation}\label{eq32}
{B_\zeta }\left( {{{\pmb{k}}_{\rm{1}}},{{\pmb{k}}_{\rm{2}}},{{\pmb{k}}_{\rm{3}}}} \right) =  - \frac{6}{5}{f_{NL}}\left[ {{P_\zeta }\left( {{{\pmb{k}}_{\rm{1}}}} \right){P_\zeta }\left( {{{\pmb{k}}_{\rm{2}}}} \right) + cyclic} \right].
\end{equation}
During our warm $k$-inflationary model, only one inflaton field is relevant, so the relation Eq. (\ref{eq27}) is reduced to
\begin{equation}\label{eq33}
\zeta \left( {t,{\pmb{x}}} \right) = {N_{\phi}}\delta\phi  + \frac{1}{2}{N_{\phi\phi}}{\left( {\delta {\phi}} \right)^2}.
\end{equation}
Thus, the general $\delta N$ part nonlinear parameter for our model can be described as
\begin{equation}\label{eq34}
- \frac{3}{5}f_{NL}^{\delta N} = \frac{1}{2}\frac{{{N_{\phi\phi}}}}{{N^2_{\phi}}}.
\end{equation}

The term $f_{NL}^{\delta N}$ is scale-independent and can be obtained by Eq. (\ref{eq34}). Inflation observations are computed at the time of the horizon crossing. Since horizon crossing occurs within the region of slow roll inflation, it is appropriate to compute the $\delta N$ part nonlinear parameter $ f_{NL}^{\delta N}$ using slow roll approximations.

Considering the conditions of the slow roll, we have
\begin{equation}\label{eq35}
	{N_\phi } =  \frac{\sigma }{{2\sqrt 3 {M_p}}}\sqrt { - K\left( \phi  \right)},
\end{equation}
and from Eq. (\ref{eq35}), we have
\begin{equation}\label{eq36}
	{N_{\phi \phi }} =  - \frac{\sigma }{{4\sqrt 3 {M_p}}}{\left[ - K\left( \phi  \right)  \right]^{ - \frac{1}{2}}}K_{\phi}.
\end{equation}
In terms of $ - \frac{3}{5}f_{NL}^{\delta N} = \frac{1}{2}\frac{{{N_{\phi \phi }}}}{{{N^2_\phi }}} $, one can obtain
\begin{equation}\label{eq37}
	 f_{NL}^{\delta N} = \frac{5}{6} \frac{{\sigma \sqrt 3 {M_p}{K_\phi }}}{{{{\left( { - K} \right)}^{\frac{3}{2}}}}} =  - \frac{5}{{12}}\sigma\epsilon,
\end{equation}
where $ \epsilon= \frac{{{K_\phi }\dot \phi }}{{HK}}$ is a slow roll parameter. Thus, we have $ f_{NL}^{\delta N} \ll 1 $.

As suggested by the warm $k$-inflation slow roll conditions, the amplitude of $\delta N $-form non-Gaussianity is not distinct in the slow roll regime, which can increase slightly accompanying the inflation for the universe. Given that the $ \delta N $ form non-Gaussianity is insufficiently large, using this part to show the whole non-Gaussianity caused by inflation is not enough and is not safe and complete, as some studies have shown \cite{Peng2016,Zhang2016}. In this case, calculating the intrinsic non-Gaussianity produced by the three-point correlation functions of the inflation field is essential.

\subsection{\label{sec32}the intrinsic part non-Gaussianity}
Compared with cold inflation, warm inflation fluctuations are generated mainly by thermal fluctuations. In warm $k$-inflation, only one scalar field plays the role of inflaton. When small perturbations are considered, the full inflaton field can be extended as $\Phi \left( {{\pmb{x}},t} \right) = \phi \left( t \right) + \delta \phi \left( {{\pmb{x}},t} \right)$, where $\delta \phi$ is the usual perturbation field surrounding the homogeneous background field $\phi \left( t \right)$.

Horizon crossing occurs inside the slow roll regime, and the observations of inflation are calculated at this time. In the warm $k$-inflation, due to the enhancement of the Hubble and thermal damped terms, the inflation evolution is overdamped inside the slow roll regime. The motion of the entire field perturbation can be explained by introducing random thermal noise $ \xi $ \cite{Peng2018},
\begin{equation}\label{eq38}
{{\cal L}_X}c_s^{ - 2}\delta {{\ddot \phi }_k}\left( t \right) + 3H\left( {{{\cal L}_X}c_s^{ - 2} + r} \right)\delta {{\dot \phi }_k}\left( t \right) + {{\cal L}_X}\frac{{{k_c^2}}}{{{a^2}}}\delta \phi_k \left( t \right) = {\xi _k},
\end{equation}
The above equation is known as the Langevin equation, and it is used to describe the interaction between a scalar field and radiation. In the above equation, $ k_c $ is the comoving wavenumber. Guaranteed by the conditions of slow roll, the inertia term $ \delta {{\ddot \phi }_k} $ is usually omitted to simplify the perturbation calculations \cite{Berera2000,Berera1996}.

To calculate $ \delta \phi  $, we expand $ \delta \phi $ to second-order $ \delta \phi  = \delta {\phi _1} + \delta {\phi _2} $, where $ \delta {\phi _1}={\cal O}\left( {\delta \phi } \right) $, and $ \delta {\phi _2}={\cal O}\left( {\delta {\phi ^2}} \right) $.
Consequently, the evolution equation of the first- and second-order perturbation field in the Fourier space can be obtained:
\begin{widetext}
\begin{equation}\label{eq39}
\frac{{d\delta {\phi _1}\left( {{\pmb{k}},t} \right)}}{{dt}} = \frac{1}{{3H\left( {6X + K + r} \right)}}\left[ { - {{\cal L}_X}{k^2}\delta {\phi _1}\left( {{\pmb{k}},t} \right) + \xi \left( {{\pmb{k}},t} \right)} \right],
\end{equation}
and
\begin{eqnarray}\label{eq40}
	\frac{{d\delta {\phi _2}\left( {{\pmb{k}},t} \right)}}{{dt}} = &&\frac{1}{{3H\left( {6X + K + r} \right)}}\left[ { - {{\cal L}_X}{k^2}\delta {\phi _2}\left( {{\pmb{k}},t} \right)} \right.
	- {k^2}{{\cal L}_{X\phi }}\int {\frac{{d{p^3}}}{{{{\left( {2\pi } \right)}^3}}}\delta {\phi _1}\left( {{\pmb{p}},t} \right)\delta {\phi _1}\left( {{\pmb{k}} - {\pmb{p}},t}  \right)}\nonumber\\
	&&\left. { - {k^2}{{\cal L}_{XX}}\int {\frac{{d{p^3}}}{{{{\left( {2\pi } \right)}^3}}}\delta {\phi _1}\left( {{\pmb{p}},t} \right)\delta {X_1}\left( {{\pmb{k}} - {\pmb{p}},t} \right)} } \right].
\end{eqnarray}

The quantity $X_1$ in the above equation can be obtained
\begin{equation}\label{eq41}
	\delta {X_1} = \dot \phi \delta \dot \phi_1  = \sqrt {2X} \frac{d}{dt}\delta {\phi _1},
\end{equation}
where $ k $ is the physical wavenumber, $ {\pmb{k}} \equiv {{\pmb{k}}_P} = \frac{{{{\pmb{k}}_{\rm{c}}}}}{a} $ ($ {{{\pmb{k}}_{\rm{c}}}} $ represents the comoving momentum, $ {{{\pmb{k}}_{\rm{p}}}} $ represents the physical momentum, and $k=|\pmb{k}|$).

By solving the evolution equations, we have

\begin{eqnarray}\label{eq42}
\delta {\phi _1}\left( {{\pmb{k}},t} \right) =&& \frac{1}{{3H\left( {6X + K + r} \right)}}\exp \left[ { - \frac{{{{\cal L}_X}{k^2}}}{{3H\left( {6X + K + r} \right)}}\left( {\tau  - {\tau _0}} \right)} \right]
\int_{{\tau _0}}^\tau  {d\tau '} \exp \left[ {\frac{{{{\cal L}_X}{k^2}}}{{3H\left( {6X + K + r} \right)}}\left( {\tau ' - {\tau _0}} \right)} \right]\xi \left( {{\pmb{k}},\tau '} \right) + \delta {\phi _1}\left( {{\pmb{k}},{\tau _0}} \right)\nonumber\\
    &&\exp \left[ { - \frac{{{{\cal L}_X}{k^2}}}{{3H\left( {6X + K + r} \right)}}\left( {\tau  - {\tau _0}} \right)} \right],
\end{eqnarray}

and
\begin{eqnarray}\label{eq43}
\delta {\phi _2}\left( {{\pmb{k}},\tau } \right) =&&\exp \left[ { - \frac{{{{\cal L}_X}{k^2}}}{{3H\left( {6X + K + r} \right)}}\left( {\tau  - {\tau _0}} \right)} \right]\int_{{\tau _0}}^\tau  {d\tau '} \exp
	\left[ {\frac{{{{\cal L}_X}{k^2}}}{{3H\left( {6X + K + r} \right)}}\left( {\tau ' - {\tau _0}} \right)} \right]\left[ {A\left( {{{k}},\tau '} \right)\int {\frac{{d{p^3}}}{{{{\left( {2\pi } \right)}^3}}}\delta {\phi _1}\left( {\pmb{p},\tau '} \right)} } \right.\nonumber\\
	&&\left. {\delta {\phi _1}\left( {{\pmb{k}} - {\pmb{p}},\tau '} \right) + B\left( {{{k}},\tau '} \right)\int\frac{{d{p^3}}}{{{{\left( {2\pi } \right)}^3}}}\delta {\phi _1}\left( {\pmb{p},\tau '} \right)\xi\left( {{\pmb{k}} - {\pmb{p}},\tau '} \right)} \right]+\delta {\phi _2}\left( {{\pmb{k}},{\tau _0}} \right)\exp \left[ { - \frac{{{{\cal L}_X}{k^2}}}{{3H\left( {6X + K + r} \right)}}\left( {\tau  - {\tau _0}} \right)} \right].
\end{eqnarray}
\end{widetext}

The parameters $ {A\left( {{\pmb{k}},\tau } \right)} $ and ${B\left( {{\pmb{k}},\tau } \right)} $ appear in above equation are

\begin{equation}\label{eq44}
	A\left( {{\pmb{k}},\tau } \right) =  - \frac{1}{{3H\left( {6X + K + r} \right)}}\left[ {{k^2}{{\cal L}_{X\phi }} + {k^2}{{\cal L}_{XX}}\frac{{\sqrt {2X} {k^2}{{\cal L}_X}}}{{3H\left( {6X + K + r} \right)}}} \right],
\end{equation}

and
\begin{equation}\label{eq45}
B\left( {{\pmb{k}},\tau } \right) = - \frac{{{{\cal L}_{XX}}{k^2}\sqrt {2X} }}{{{{\left[ {3H\left( {6X + K + r} \right)} \right]}^2}}}.
\end{equation}

This paper defines the parameter $ \tau \left( \phi  \right) = \frac{{3H\left( {{{\cal L}_X}c_s^{ - 2} + r} \right)}}{{{{\cal L}_X}{k^2}}} $ based on Eq. (\ref{eq42}) to characterize the efficiency of the thermalizing process. It is found a larger $ k $ indicates a faster relaxation rate. If the $ k $ of one of the fields $ \Phi(\mathbf{x},t) $ is large enough to relax within a Hubble time, the mode can be thermal. When the physical wave number $ k_p $ of the corresponding $\Phi(\mathbf{x},t)$ mode is smaller than the freeze-out physical wave number $ k_F $, it is no longer affected by thermal noise $ {\xi _k} $ during a Hubble time. Based on the condition $ \frac{{3H\left( {{{\cal L}_X}c_s^{ - 2} + r} \right)}}{{{{\cal L}_X}k^2}} = \frac{1}{H} $, the freeze-out momentum $ k_F $ could be given by
\begin{equation}\label{eq46}
{k_F} = \sqrt {\frac{{3{H^2}}}{{c_s^2}}\left( {1 + \frac{{rc_s^2}}{{{{\cal L}_X}}}} \right)}.
\end{equation}

As previously stated, the first-order inflaton perturbation $ \delta {\phi _1} $ is a pure Gaussian field, and their bispectrum vanishes due to their statistical stochastic features. To calculate non-Gaussianity, the bispectrum resulting from two first-order and one second-order fluctuations should have the highest order. Then we have
\begin{widetext}
\begin{eqnarray}\label{eq47}
		&&\left\langle {\delta \phi \left( {{{\pmb{k}}_{\rm{1}}},\tau } \right)\delta \phi \left( {{{\pmb{k}}_{\rm{2}}},\tau } \right)\delta \phi \left( {{{\pmb{k}}_{\rm{3}}},\tau } \right)} \right\rangle\nonumber\\
&&=\exp \left[ { - \frac{{{{\cal L}_X}{k^2}}}{{3H\left( {6X + K + r} \right)}}\left( {\tau  - {\tau _0}} \right)} \right]\int_{{\tau _0}}^\tau  {d\tau '}
	 \exp \left[ {\frac{{{{\cal L}_X}{k^2}}}{{3H\left( {6X + K + r} \right)}}\left( {\tau ' - {\tau _0}} \right)} \right]\left[ {A\left( {{{k}},\tau '} \right)\int {\frac{{d{p^3}}}{{{{\left( {2\pi } \right)}^3}}}} } \right.\nonumber\\
	&& \left\langle {\delta {\phi _1}\left( {{{\pmb{k}}_1},\tau '} \right)\delta {\phi _1}\left( {{{\pmb{k}}_2},\tau '} \right)\delta {\phi _1}\left( {{\pmb{p}},\tau '} \right)\delta {\phi _1}\left( {{{\pmb{k}}_3} - {\pmb{p}},\tau '} \right)} \right\rangle
	+ B\left( {{{k}},\tau '} \right)\int\frac{{d{p^3}}}{{{{\left( {2\pi } \right)}^3}}}\left\langle {\delta {\phi _1}\left( {{{\pmb{k}}_1},\tau } \right)\delta {\phi _1}\left( {{{\pmb{k}}_2},\tau } \right)\delta {\phi _1}\left( {{\pmb{p}},\tau '} \right)} \right.\nonumber\\
	&& \left. {\left. {\xi \left( {{\pmb{k}_3} - \pmb{p},\tau '} \right)} \right\rangle } \right] + \exp \left[ { - \frac{{{L_X}{k^2}}}{{3H\left( {6X + K + r} \right)}}\left( {\tau  - {\tau _0}} \right)} \right]\left\langle {\delta {\phi _1}\left( {{{\pmb{k}}_1},\tau } \right)\delta {\phi _1}\left( {{{\pmb{k}}_2},\tau } \right)\delta {\phi _2}\left( {{{\pmb{k}}_3},{\tau _0}} \right)} \right\rangle  + \left( {{{\pmb{k}}_1} \leftrightarrow {{\pmb{k}}_3}} \right)
+ \left( {{{\pmb{k}}_2} \leftrightarrow {{\pmb{k}}_3}} \right).
\end{eqnarray}
\end{widetext}

The bispectrum amplitude is determined when the cosmic scale departs the horizon. There are about 60 e-folds until the end of the inflation, and $ {{{\pmb{k}}_1}} $, $ {{{\pmb{k}}_2}} $, and $ {{{\pmb{k}}_3}}$ are crossing the horizon all within a few e-folds. $ k_F> H $ is deduced from the expression of $ k_F $ in the warm k-inflationary model. This indicates the correlations in our model, known as the thermalized correlations, that should be calculated at the crossing of the Hubble horizon $ k= H $, are determined at an earlier freeze-out period $k= k_F$ \cite{Berera2000,Bass1999,Hecke1998}. Thus, the duration between corrections can be calculated by
\begin{equation}\label{eq48}
	\Delta {t_F} = {t_H} - {t_F} \simeq \frac{1}{H}\ln \left( {\frac{{{k_F}}}{H}} \right).
\end{equation}
The bispectrum can then be reduced to
\begin{eqnarray}\label{eq49}
	&&\left\langle {\delta \phi \left( {{{\pmb{k}}_{\pmb{1}}},t} \right)\delta \phi \left( {{{\pmb{k}}_{\rm{2}}},t} \right)\delta \phi \left( {{{\pmb{k}}_{\rm{3}}},t} \right)} \right\rangle  \simeq 2A\left( {{k_F},t_F} \right)\Delta {t_F}\nonumber\\
	&&\times \left[ {\frac{{d{p^3}}}{{{{\left( {2\pi } \right)}^3}}}\left\langle {\delta {\phi _1}\left( {{{\pmb{k}}_1},t} \right)\delta {\phi _1}\left( {{{\pmb{k}}_1},p} \right)} \right\rangle \left\langle {\delta {\phi _1}\left( {{\pmb{k}_2},t} \right)} \right.} \right.\nonumber\\
	&&\left. {\left. {\delta {\phi _1}\left( {{{\pmb{k}}_3} - p,t} \right)} \right\rangle  + \left( {{{\pmb{k}}_1} \leftrightarrow {{\pmb{k}}_3}} \right) + \left( {{{\pmb{k}}_2} \leftrightarrow {{\pmb{k}}_3}} \right)} \right].
\end{eqnarray}

According to Eqs. (\ref{eq31}), (\ref{eq32}), (\ref{eq49}), and the relation $\zeta=\frac{H}{\dot\phi}\delta\phi$, the intrinsic non-Gaussian nonlinear parameter can be obtained,
\begin{eqnarray}\label{eq50}
f_{NL}^{int}=&-&\frac{5}{6}\frac{{\dot \phi }}{H}2A\left( {{k_F},{t_F}} \right)\Delta {t_F}\nonumber\\ =&&\frac{5}{3}\ln \sqrt {\frac{3}{{c_s^2}}\left( {1 + \frac{{rc_s^2}}{{{{\cal L}_X}}}} \right)}\nonumber\\
&&\left[ {\frac{{\sqrt {2X} k_F^2{{\cal L}_{X\phi }}}}{{3{H^3}\left( {6X + K + r} \right)}} + \frac{{2 k_F^4{{\cal L}_{XX}}{{\cal L}_X}X}}{{9{H^4}{{\left( {6X + K + r} \right)}^2}}}} \right]\nonumber\\
=&&\underbrace {\frac{5}{3}\ln \sqrt {\left( {\frac{3}{{c_s^2}} + \frac{{3r}}{{{{\cal L}_X}}}} \right)} \frac{K}{{K + 2X}}\left( {1 + \frac{{rc_s^2}}{{{{\cal L}_X}}}} \right)\epsilon}_{term1} \nonumber\\ &+& \underbrace {\frac{5}{3}\ln \sqrt {\left( {\frac{3}{{c_s^2}} + \frac{{3r}}{{{{\cal L}_X}}}} \right)} \frac{{4X\left( {2X + K} \right)}}{{{{\left( {6X + K + r} \right)}^2}}}{{\left( {\frac{1}{{c_s^2}} + \frac{r}{{{{\cal L}_X}}}} \right)}^2}}_{term2}.
\end{eqnarray}
By exploiting conditions of slow roll in warm $k$-inflation, we have

\begin{equation}\label{eq51}
term{\rm{ }}1 = \frac{5}{3}\ln \sqrt {\left( {\frac{3}{{c_s^2}} + \frac{{3r}}{{{{\cal L}_X}}}} \right)} \frac{K}{{K + 2X}}\left( {1 + \frac{{rc_s^2}}{{{{\cal L}_X}}}} \right)\epsilon \ll 1.
\end{equation}

From this, it can be concluded that the second term dominates the intrinsic non-Gaussian nonlinear parameter, which can be obtained as follows:

\begin{equation}\label{eq52}
	f_{NL}^{int} \simeq \frac{5}{3}\ln \sqrt {\left( {\frac{3}{{c_s^2}} + \frac{{3r}}{{{{\cal L}_X}}}} \right)} \frac{{4X\left( {2X + K} \right)}}{{{{\left( {6X + K + r} \right)}^2}}}{\left( {\frac{1}{{c_s^2}} + \frac{r}{{{{\cal L}_X}}}} \right)^2}.
\end{equation}
From this equation, a small inflaton sound speed can significantly increase the amount of intrinsic non-Gaussianity, and strong thermal dissipation can also increase the proportion of intrinsic non-Gaussianity in warm $k$-inflation.

\subsection{\label{sec33}discussions of result and parameters restriction of the model}

According to the above-mentioned field evolution equation, there are two dissipation terms, namely, Hubble dissipation $3H\left( {{{\dot \phi }^2} + K} \right)\dot \phi $ and thermal dissipation $\Gamma \dot \phi $. Thus, we have ${{\cal L}_X} = K + 2X = K + {{\dot \phi }^2}$, and $r \gg {{\cal L}_X}$ indicates thermal effects dominate. This paper compares two portions of the non-Gaussianity based on the conclusions reached. $ f_{NL}^{\delta N} $ is represented by the polymerization of the redefined slow roll parameters. So, $ f_{NL}^{\delta N} $ should be far smaller than 1 in the slow roll inflationary regime, while the intrinsic part $ f_{NL}^{int} $ is much more than 1 if the sound speed of inflaton field is small enough. Since the noncanonical effect is strong, it can be concluded that the intrinsic component of non-Gaussianity is the main component.

The entire nonlinear parameter could be estimated using the nonlinear parameter that we have calculated in two parts. That is,
\begin{eqnarray}\label{eq53}
 f_{NL}^{int}  = &-& \frac{5}{{12}}\sigma \epsilon  + \frac{5}{3}\ln \sqrt {\left( {\frac{3}{{c_s^2}} + \frac{{3r}}{{{{\cal L}_X}}}} \right)} \frac{K}{{K + 2X}}\left( {1 + \frac{{rc_s^2}}{{{{\cal L}_X}}}} \right)\epsilon \nonumber\\
 &+& \frac{5}{3}\ln \sqrt {\left( {\frac{3}{{c_s^2}} + \frac{{3r}}{{{{\cal L}_X}}}} \right)} \frac{{4X\left( {2X + K} \right)}}{{{{\left( {6X + K + r} \right)}^2}}}{\left( {\frac{1}{{c_s^2}} + \frac{r}{{{{\cal L}_X}}}} \right)^2}\nonumber\\
 &\simeq &{\left( {\frac{1}{{c_s^2}} + \frac{r}{{{{\cal L}_X}}}} \right)^2}.
\end{eqnarray}

The result obtained above indicates that in the thermal effect dominated regime $ r \gg 1 $, $ {f_{NL}} \sim {\left( {\frac{r}{{{{\cal L}_X}}}} \right)^2} $. Consequently, when thermal effects dominate in the universe, the non-Gaussianity is obvious. In the weak dissipative regime $ r \ll 1 $, $ {f_{NL}} \sim \frac{1}{{c_s^4}} $, as determined mainly by the sound speed. Thus, if the universe is dominated by noncanonical effects, the non-Gaussianity is more obvious. We can see that both the thermal and noncanonical effects can enhance the magnitude of non-Gaussianity. Due to observational limitations $ f_{NL}\sim{\cal O}\left( {{{10}^2}} \right) $ \cite{Bartolo2016,PLANCK2}, the speed of sound $c_s$ must not be too small, and the dissipative strength parameter $r$ is required to be not extremely large. If the warm $k$-inflationary model can fit the observations well, neither the noncanonical effect nor the thermal effect should be too strong.
\section{\label{sec4}conclusions and discussions}

This paper investigates the entire primordial non-Gaussianity produced by warm $k$-inflation. The essential equations of warm $k$-inflation are presented, such as the motion equation, e-folds, slow roll equations, as well as slow roll conditions. This paper emphasizes the key problem: non-Gaussianity resulted from warm $k$-inflation. The nonlinear parameter is usually used to quantify the degree of non-Gaussianity, and it consists of two components: the intrinsic part $ f_{NL}^{int} $ and the $ {\delta N} $ part $ f_{NL}^{\delta N} $. The first component covers the impact of the three-point correlation, i.e., intrinsic non-Gaussianity of the inflaton field. The second component is determined by a four-point correlation with inflaton perturbations. The original non-Gaussianity in warm $k$-inflation can be fully captured by these two components.

The formalism of $\delta N$ is introduced and used to calculate the $ {\delta N} $ part non-Gaussianity. It is concluded from the obtained results that $ f_{NL}^{\delta N} $ is defined as the linear combination of the redefined slow roll parameters. So, in slow roll inflation, $ f_{NL}^{\delta N}  $ is a first-order small quantity, and it indicates that the $ {\delta N} $ part non-Gaussianity for warm $k$-inflation is not significant. However, the situation is not the same for calculating intrinsic non-Gaussianity. The intrinsic non-Gaussianity is principally driven by the sound speed and dissipation strength parameters, and it is produced by three-point correlations in the inflation. Throughout the entire non-Gaussianity in warm $k$-inflation, it is observed that$ f_{NL}^{int} $ dominates the $ f_{NL}^{\delta N} $ part, and sound speed plays the most important role in non-Gaussianity of our model, and thermal dissipation effects also contribute to non-Gaussianity.

\section{Acknowledgments}
This work was supported by the Shandong Provincial Natural Science Foundation, China (Grant No. ZR20MA037, ZR2021QA037 and ZR2022JQ04), National Natural Science Foundation of China (Grant No. 12205158 No. 11975132 and No. 11605100), the Henan Provincial Natural Science Foundation of China under Grant No. 232300421351 and the Talent Introduction Fund (Grant No. 2020BS035) at Henan University of Technology.

\end{document}